\documentclass{Interspeech}

\interspeechcameraready

\title{REWIND: Speech Time Reversal for Enhancing Speaker Representations in Diffusion-based Voice Conversion}

\author[affiliation={1}]{Ishan D.}{Biyani}
\author[affiliation={1}]{Nirmesh J.}{Shah}
\author[affiliation={1}]{Ashishkumar P.}{Gudmalwar}
\author[affiliation={1}]{Pankaj}{Wasnik}
\author[affiliation={2}]{Rajiv R.}{Shah}

\affiliation{Media Analysis Group}{Sony Research India}{Bangalore}
\affiliation{Indraprastha Institute of Information Technology (IIIT)}{Delhi}{India}
\email{\{ishan.biyani@sony.com,nirmesh.shah, ashish.gudmalwar1,pankaj.wasnik\}@sony.com; rajivratn@iiitd.ac.in}

\usepackage{comment}
\usepackage{multirow}
\usepackage{cite}
\begin{document}
\maketitle
\begin{abstract}
Speech time reversal refers to the process of reversing the entire speech signal in time, causing it to play backward. Such signals are completely unintelligible since the fundamental structures of phonemes and syllables are destroyed. However, they still retain tonal patterns that enable perceptual speaker identification despite losing linguistic content. In this paper, we propose leveraging speaker representations learned from time reversed speech as an augmentation strategy to enhance speaker representation. Notably, speaker and language disentanglement in voice conversion (VC) is essential to accurately preserve a speaker's unique vocal traits while minimizing interference from linguistic content. The effectiveness of the proposed approach is evaluated in the context of state-of-the-art diffusion-based VC models. Experimental results indicate that the proposed approach significantly improves speaker similarity-related scores while maintaining high speech quality.
\end{abstract}
\section{Introduction}
Voice Conversion (VC) is the process of altering the perceived speaker identity in a speech signal from a source speaker to that of a target speaker while retaining the original linguistic content \cite{mohammadi2017overview}. This technology has significant applications in the entertainment industry, where it is used for tasks such as dubbing, character voice transformation during post production in films, video games, advertisements, and animations \cite{brannon2023dubbing,hu2021neural,sahipjohn2024dubwise,mhaskar-etal-2024-isometric,gudmalwar2024vecl}. In many real-world scenarios, voice conversion must be achieved with minimal available data, often relying on just one or even no prior recordings of the target voice, making one-shot and zero-shot VC approaches critical \cite{wang2023lm,li2024sef,yao2024stablevc,hwang2022stylevc,shah2023nonparallel}. These approaches enable models to adapt to new, unseen voices providing the necessary flexibility. Their importance lies in the ability to rapidly generate high-quality, natural-sounding speech for any target voice, thereby enhancing creative possibilities and streamlining production workflows in the entertainment industry.\\
\indent In the context of VC, capturing speaker identity is crucial and is typically achieved through speaker encoders. These encoders can be either pre-trained for speaker classification tasks utilizing large datasets to learn robust, speaker specific embeddings\cite{chen2024controlvc} or they can be trained jointly with the VC model directly\cite{Seki2023jsv,Guo2023joint}. The resulting speaker embeddings should capture key vocal attributes such as timbre, pitch, and prosody, which are essential for characterizing a speaker's unique identity \cite{wang2017does}. However, these embeddings often suffer from linguistic interference, where language related features entangled with the speaker specific characteristics. To address this, language disentanglement strategies are often employed during training\cite{Mun2022dis, Nam2022dis, Tu2023con}, ensuring that the learned embeddings remain as language independent as possible. Finally, these embeddings are used to condition the decoder within the VC framework to generate speech that perceptually reflects the target speaker's identity while preserving the original linguistic content.\\
\indent In this paper, we propose the concept of speech time reversal (STR) as a novel data augmentation strategy to enhance speaker representations. Speech time reversal involves inverting the entire speech signal along the time axis so that it plays backward. Previous studies have utilized short-time speech reversal for noise robust automatic speech recognition (ASR) task \cite{chao2021tenet} because it alters the signal in a controlled manner while still preserving most of the linguistic content. This controlled modification helps models become robust to acoustic variations and distortions without losing the key phonetic cues needed for these tasks. In contrast, complete speech time reversal has been largely overlooked in the literature. Although speech time reversal process makes the signal completely unintelligible by destroying the core phonemic and syllabic structures, it preserves rhythmic and tonal patterns that enable perceptual speaker identification\cite{saberi1999cognitive}. We initially conducted perceptual studies to determine whether speaker identity is retained in time-reversed speech signals. Based on these findings, we extended our hypothesis to voice conversion by learning speaker embeddings from time reversed speech and fusing them with conventional speaker embeddings. The combined speaker embeddings were subsequently used to condition the diffusion-based decoder in the VC framework. We conducted several subjective and objective evaluations on the LibriTTS\cite{zen2019libritts} and VCTK\cite{yamagishi2019cstr} databases to demonstrate the effectiveness of the proposed strategy against state-of-the-arts (SOTA) VC methods.
\section{Related work}
Early voice conversion (VC) systems relied on traditional statistical models \cite{mohammadi2017overview}, such as Gaussian Mixture Models (GMMs), to establish a mapping between the acoustic features of source and target speakers\cite{stylianou1998continuous}. These early approaches \cite{mohammadi2017overview}, while pioneering, struggled with the high-dimensional and nonlinear aspects of speech. With the advent of deep learning, research in VC evolved rapidly, transitioning through deep neural networks (DNNs)\cite{chen2014voice}, sequence-to-sequence \cite{zhang2019sequence}, variational autoencoders (VAEs)\cite{hsu2016voice}, and generative adversarial networks (GANs)\cite{zhang2020gazev} to achieve more natural and expressive speech conversion. Recently, diffusion based models have gained traction due to their robust training dynamics and iterative denoising process\cite{popov2021diffusion, choi2023diffhvc, choi2024dddm, gudmalwar2025emoreg}, which allows for the generation of high quality speech even in challenging low-data scenarios. \\
\indent Controlling speaker identity in VC has remained a central challenge, particularly in zero-shot and one-shot settings \cite{wang2023lm,li2024sef,yao2024stablevc,hwang2022stylevc}. To address this, researchers have integrated speaker embedding networks into the conversion framework. These networks, often pretrained on large-scale speaker recognition tasks to produce robust embeddings such as d-vectors or x-vectors, capture essential vocal attributes like timbre, pitch, and prosody \cite{chen2024controlvc}. Some approaches use a two-stage training process where the speaker encoder is pretrained independently and later fused with the VC model as a conditioning mechanism. Other methods opt for joint training\cite{qian2019autovc}, ensuring that the speaker embeddings are directly optimized for the conversion task. Additionally, techniques such as adaptive normalization and attention-based fusion have been employed to further refine the control over the target speaker's identity, significantly enhancing speaker similarity and overall conversion quality.
\section{Proposed Approach}
\subsection{Proposed Speech Time Reversal (STR) Strategy}
\begin{figure}[h]
\vspace{-0.2cm}
  \centering
  \includegraphics[width=0.95\linewidth]{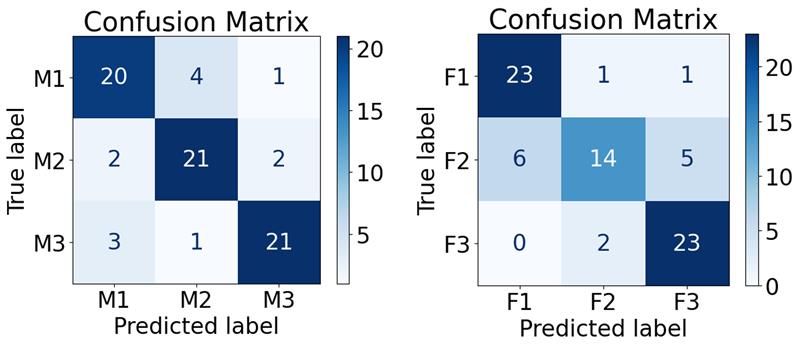}
  \vspace{-0.3cm}
  \caption{Confusion matrices for the perceptual study of speaker identification from the time reversed speech. Here, M1, M2, M3 and F1, F2, F3 represents three different Male and Female speakers, respectively.}
  \label{fig:CM}
  \vspace{-0.6cm}
\end{figure}

\begin{figure}[t]
  \centering
  \includegraphics[width=\linewidth]{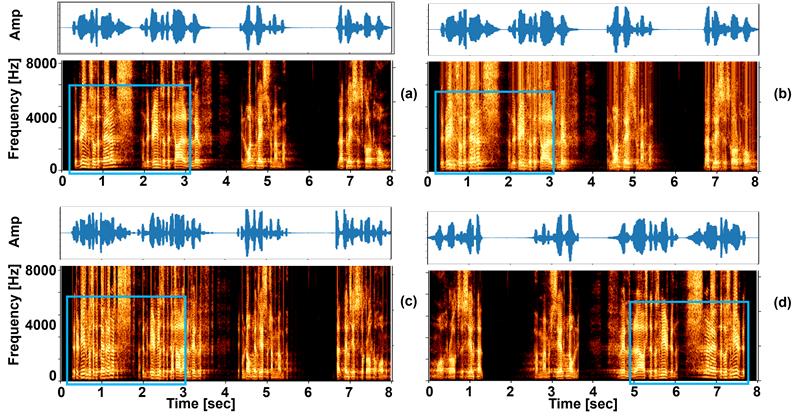}
  \vspace{-0.5cm}
  \caption{Spectrographic visualization of (a) original speech (b) 20 ms short-time, (c) 100 ms short-time speech reversal, and (d) complete speech time reversal.}
  \label{fig2:spec-analysis}
  \vspace{-0.7cm}
\end{figure}

Let $x(t)$ denotes the original signal, then the time reversed speech signal is obtained by flipping $x(t)$ along time axis, i.e., $x(l-t)$, where  $t$ $\epsilon$  $[0,l]$. To assess the preservation of speaker identity in the time reversed speech, we designed a controlled perceptual study. In particular, we asked subjects to listen to an original reference sample and then to identify the correct speaker from a set of time-reversed speech signals. We utilized English speech data from total six speakers consisting of three males and three females. Total 25 participants participated in the perceptual study. The goal of this study was to determine whether the unique vocal attributes of a speaker, such as timbre and intonation, remain perceptually recognizable after the complete speech time reversal process. By comparing the reference recording with the altered samples, we aimed to evaluate the effectiveness of time reversal in retaining speaker identity, despite the loss of intelligible linguistic content. In particular, subjects could identify the correct speaker with 80.3\% accuracy from the time reversed speech signals. Corresponding confusion matrices for the percpetual study is shown in Figure \ref{fig:CM}.
\begin{table}[h]
\caption{WER and speaker similarity (SS) score for different speech time reversal strategies.}
\vspace{-0.3cm}
\setlength{\tabcolsep}{1pt}
\resizebox{\columnwidth}{!}{%
\begin{tabular}{cccccccc}
\hline
\multirow{2}{*}{} & \multicolumn{6}{c}{Short Time}                                                          & Complete Time\\ \cline{2-8} 
                   & \multicolumn{1}{c}{10ms} & \multicolumn{1}{c}{20ms} & \multicolumn{1}{c}{50ms} & \multicolumn{1}{c}{100ms} & \multicolumn{1}{c}{200ms} & 500ms & Full  length  \\ \hline
SS Score      & \multicolumn{1}{c}{0.89} & \multicolumn{1}{c}{0.90} & \multicolumn{1}{c}{0.86} & \multicolumn{1}{c}{0.91} & \multicolumn{1}{c}{0.94} & 0.94 & \textbf{0.96}      \\ 
WER                 & \multicolumn{1}{c}{0\%} & \multicolumn{1}{c}{0\%} & \multicolumn{1}{c}{28\%} & \multicolumn{1}{c}{100\%} & \multicolumn{1}{c}{100\%} & 100\% & 100\%      \\ \hline
\end{tabular}
}
\vspace{-0.2cm}
\end{table}
\\ \indent Additionally, to compare the proposed speech time reversal strategy with the short-time speech reversal approach, we analyzed the spectrographic outputs from both methods. Figure \ref{fig2:spec-analysis} presents the spectrographic visualization of (a) original speech, (b) 20 ms short-time speech reversal and (c) 100 ms short-time speech reversal strategy and (d) complete speech time reversal. Our analysis revealed that, in the case of complete speech reversal, the harmonic structures are prominently visible, which strongly indicates the retention of speaker-specific information. The clear presence of these harmonic patterns suggests that even though the reversed speech is rendered unintelligible, it preserves critical acoustic cues such as timbre and pitch that are unique to the speaker. To further substantiate these findings, we calculated Word Error Rates (WER) and speaker similarity scores based on cosine similarity metric as shown in Table 1. The evaluation showed that the complete reversal method not only maintains a higher WER but also yields higher speaker similarity scores compared to the short-time speech reversal strategy. These results, aligning with \cite{saberi1999cognitive}, confirm that complete speech reversal is more effective in preserving speaker identity with no linguistic interference, making it a promising approach for VC application.
\subsection{Proposed Diffusion-based Voice Conversion}
\begin{figure*}
  \centering
  \includegraphics[width=0.8\linewidth]{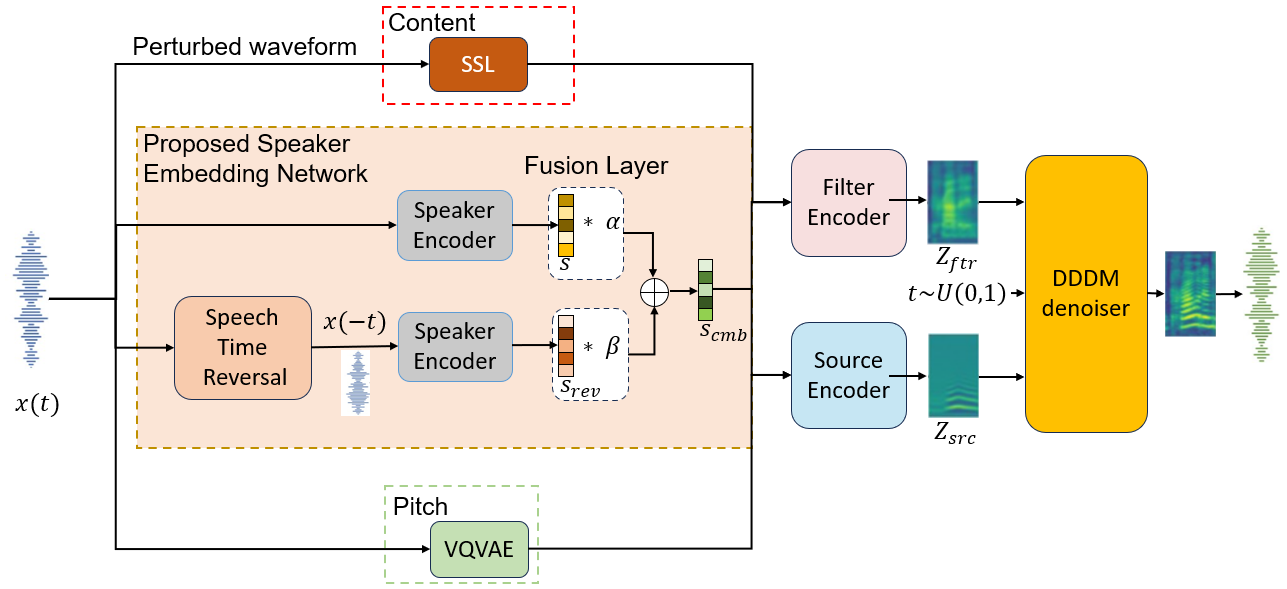}
  \vspace{-0.3cm}
  \caption{Blockdiagram of the propose approach in Diffusion-based Voice Conversion}
  \vspace{-0.7cm}
  \label{fig:block-diagram}
\end{figure*}
Denoising diffusion models have emerged as a powerful class of generative models, significantly advancing task of voice conversion \cite{choi2023diffhvc,choi2024dddm,popov2021diffusion}. These models work in two stages: a forward diffusion process that incrementally adds noise to the data and a reverse denoising process that gradually recovers the original signal. Traditional diffusion models use a discrete-time Markov chain \cite{ho2020denoising}, while score-based generative modeling employs continuous-time stochastic differential equations \cite{Song2021score} for greater flexibility and improved sample quality. In particular, decoupled denoising diffusion models (DDDMs) in DDDM-VC leverage multiple disentangled denoisers to control speaking style attributes effectively \cite{choi2024dddm}. Our proposed STR strategy is implemented using DDDM-VC as the backbone architecture.

DDDM-VC features both a source-filter encoder and decoder and utilizes self-supervised speech representations to disentangle various aspects of the speech signal. It employs a prior mixup technique to enhance robustness across different conversion scenarios. The model uses three types of representations for speech disentanglement: (1) Content Representation, which extracts phonetic content using models like XLS-R\cite{babu2021xls} from Wav2Vec 2.0; (2) Pitch Representation, where the YAPPT algorithm extracts the fundamental frequency ($F_0$)—normalized per utterance and processed using VQ-VAE—to capture intonation; and (3) Speaker Representation, where a speaker embedding network extracts speaker representation from the target Mel-spectrogram. These features are averaged at the utterence level and integrated across the network to enable robust zero-shot voice style transfer. To further enhance the speaker representation, we propose combining the conventional speaker representation with one derived from the speech time reversal strategy. The two embeddings are then integrated using a weighted fusion layer (as shown in Figure \ref{fig:block-diagram}). The combined speaker representation ($S_{cmb}$) is defined as follows:
\begin{align}
  S_{cmb}=\alpha * S + \beta * S_{rev}
\end{align}
where $S$ and $S_{rev}$ denotes the speaker embeddings learned from the original and time-reversed speech, respectively. The weighting coefficients $\alpha$ and $\beta$ are selected from the range $[0, 1]$.

\indent Finally model uses source-filter theory \cite{fant1970acoustic} for speech resynthesis. While the source encoder processes pitch and speaker information, the filter encoder deals with content and speaker information. A data-driven prior in the diffusion process can successfully direct the beginning point of the reverse process. This is improved by using fully reconstructed source and filter Mel-spectrograms, \(Z_{src}\) and \(Z_{ftr}\), respectively, to provide a more thorough prior. These are regularized against the target Mel-spectrogram \(X_{mel}\) with the following loss function:
\begin{align}
\mathcal{L}_{rec} &= \lvert\lvert X_{mel} - (Z_{src} + Z_{ftr}) \rvert\rvert_1,
\end{align}
where $Z_{src} = E_{src}(pitch, s)$, $Z_{ftr} = E_{ftr}(content, S_{cmb})$. 

It is also important to note that the disentangled source and filter Mel-spectrograms are converted using different speaker representations. The entire model is jointly trained end-to-end, including the style encoder, source-filter encoder, and decoder, without pre-trained XLS-R and $F_0$ VQ-VAE.
\section{Experimental Analysis}
\subsection{Experimental Setup}
We trained our model on LibriTTS (1,100 speakers, 10 held out) and evaluated it on the held-out speakers as well as the VCTK dataset. For self-supervised speech representations, audio was downsampled from 24 kHz to 16 kHz and fed into XLS-R (0.3B). We generated 80-D log-scale Mel spectrograms for the target voice and speaker encoder using a hop size of 320 a window size of 1280, and a 1280 point Fourier transform in order to match the time frames between the self-supervised representation and the Mel spectrogram without interpolation. We trained DDDM-VC using the AdamW optimizer (Loshchilov \& Hutter, 2019) with \(\beta_1 = 0.8, \beta_2 = 0.99\) and a weight decay of \(\lambda = 0.01\). The learning rate started at \(5 \times 10^{-5}\) and decayed according to a schedule of \(0.999^{\frac{1}{8}}\). Training was conducted on two NVIDIA L40 GPUs (32GB each) for 1000 epochs with a batch size of 64, taking approximately 2 days. The proposed Speaking Style Encoder uses two multi-head attention layers and a 0.1 dropout rate to process log-scale Mel-spectrograms from the input waveform with an 80 dimensional input, extracting speaker embeddings of size 256-D. Similarly, reverse speaker embeddings are extracted from the time-reversed waveform. Both embeddings are fused through a weighted fusion layer to produce the final speaker representation. We employed HiFi-GAN V1 \cite{Kong2020hifigan} for waveform synthesis, substituting multi-scale STFT-based discriminators (MS-STFTD) from EnCodec \cite{Defossez2022compression} to improve vocoder quality. The Encoder and Decoder have 6.36M and 60M parameters, respectively. 
\subsection{SOTA Baseline VC Methods}
The following is a list of the SOTA VC baseline methods used to validate the effectiveness of our proposed strategy. \\ \textbf{Diff-VC\cite{popov2021diffusion}:} A diffusion-based VC baseline that iteratively refines noisy speech signals through reverse diffusion, producing high-fidelity conversions with strong speaker adaptation.\\
\textbf{Diff-HierVC\cite{choi2023diffhvc}:} A hierarchical VC system that uses dual diffusion models DiffVoice for speech quality and DiffPitch for target $F_0$ generation along with a source-filter encoder and masked precursor to enhance speaker adaptation and pitch accuracy.\\
\textbf{LM-VC\cite{wang2023lm}:} A two-stage language modeling approach for zero-shot VC that employs a masked prefix language model and window-attentive external LM to disentangle speaker information and recover masked content.\\
\textbf{SEF-VC\cite{li2024sef}:} A speaker embedding-free VC model that leverages position-agnostic cross-attention to directly model and integrate speaker timbre from reference speech, enhancing adaptability and stability. \\
\textbf{StyleVC\cite{hwang2022stylevc}:} A VC model that mitigates training-inference mismatches using adversarial training, pitch prediction, and style generalization with unpaired samples to improve acoustic quality and speaker adaptation. \\
\textbf{StableVC\cite{yao2024stablevc}:} A style controllable zero-shot VC system that disentangles linguistic content, timbre, and style using a conditional flow matching module with dual attention, ensuring fast and high-quality synthesis. \\
\textbf{VALLE VC\cite{wang2023val}:} A model that utilizes a large scale generative speech framework and neural audio codecs to achieve expressive, high-fidelity voice conversion while preserving target speaker identity.
\subsection{Experimental Results}
The voice conversion performance is evaluated in the following section by comparing the proposed approach with baseline and state-of-the-art approaches. Demo samples are available at URL \footnote{\url{https://research.sri-media-analysis.com/interspeech25-rewind-vc/}}. Since improving speaker representation is the main goal, the speaker similarity score serves as the primary objective metric for evaluating effectiveness. This metric helps in determining how effectively the proposed method improve speaker characteristics during the voice conversion task. Table 2 provides a summary of the results, encompassing both subjective and objective assessments.
Speaker embeddings are extracted from both generated and reference audio samples using Resemblyzer\footnote{\url{https://github.com/resemble-ai/Resemblyzer}} in order to calculate the Objective Speaker Similarity Score. The 256-dimensional summary vector that Resemblyzer produces captures the characteristics of the speaker. The similarity between the generated and reference speech is then measured by computing the cosine distance between their respective embeddings. The results presented in Table 2 indicate an improvement in speaker similarity metrics for diffusion-based voice conversion baselines, particularly in the cases of DiffHierVC and DDDM-VC. Additionally, DiffVC demonstrates comparable performance. Overall the results shows an improvement of 4.16\% using proposed approach with respect to baseline highlighting its effectiveness.

Another objective evaluation metric that was used was the AutoPCP score, which measures the prosody-level similarity across speech samples using a model-based estimator of Prosody Conversion Precision (PCP) \cite{barrault2023seamless}. Better prosodic alignment is reflected in a higher AutoPCP score. The findings show that, in comparison to the baseline techniques, the suggested strategy often produces either improvements or comparable performance in AutoPCP scores. We used two Mean Opinion Score (MOS) predictions methods to evaluate MOS objectively. The first, WV-MOS, is an absolute objective speech quality measure based on a fine-tuned Wave2Vec 2.0 model for direct MOS score prediction. Since winning several categories in the 2022 VoiceMOS challenge, the second, UTMOS, is an ensemble-based MOS prediction system that has been widely used to evaluate the top TTS systems. By evaluating audio quality on a scale of 1 to 5, with 5 denoting the highest quality and 1 the lowest, UTMOS determines the mean opinion score (MOS) for a given speech sample. The proposed approach shows improvement of 0.5\% to 5.5\% in both metrics compared to baseline methods.

For subjective evaluation, the MUSHRA-based MOS metric was used to assess speech similarity. A group of 25 participants, all between the ages of 25 and 35, having no known hearing impairments, evaluated the samples on a scale of 0 to 100, where 100 is the greatest quality, depending on how similar the speaker was to the reference. Statistical significance is computed via $95\%$ confidence intervals in subjective evaluations, with MOE ranging from $\pm 1.66$ to $\pm 2.96$. According to the evaluation's findings, the proposed approach outperforms baseline techniques by $2.67\%$ w.r.t. average gain in subjective speaker similarity score.
\begin{table}[h]
\caption{Subjective and objective evaluation analysis}
\vspace{-0.3cm}
\setlength{\tabcolsep}{1pt}
\resizebox{\columnwidth}{!}{%
\begin{tabular}{c|c|c|c|c|c} \hline
          & \begin{tabular}[c]{@{}c@{}}Subjective\\ SS Score\end{tabular} & \begin{tabular}[c]{@{}c@{}}Objective \\ SS Score\end{tabular} & AutoPCP & WV-MOS & UT-MOS \\  \hline
\multicolumn{6}{c}{SOTA VC Baselines}                                                                     \\  \hline 
LMVC        & $51.28 \pm 2.48$                             & 0.74                             & 2.75  & 3.32  & 2.9  \\
SEFVC       & $43.17 \pm 1.84$                             & 0.71                             & 2.7   & 3.3  & 2.47  \\
StyleVC      & $51.97 \pm 2.91$                             & 0.76                             & 2.93  & 3.85  & 3.75  \\
StableVC      & $62.11 \pm 2.26$                             & 0.67                             & 2.89  & 4.21  & 3.78  \\
VALLE-VC      & $55.61 \pm 1.83$                             & 0.72                             & 2.87  & 3.96  & 3.63  \\  \hline 
\multicolumn{6}{c}{Diffusion-based VC baselines}                                                                \\  \hline 
DiffVC       & $50.12 \pm 1.90$                             & \textbf{0.75 }                            & \textbf{2.45}  & \textbf{3.57}  & 3.4  \\
Diff VC+Ours    & \textbf{$53.42 \pm 2.09$}                            & 0.71                             & 2.43  & 3.56  & \textbf{3.46}  \\
DiffHierVC     & \textbf{$75.76 \pm 2.96$}                            & 0.76                             & 2.88  & 3.99  & 3.68  \\
DiffHier-VC + Ours & $75.72 \pm 1.66$                        & \textbf{0.8}                 & \textbf{2.89}  & 3.99  & \textbf{3.83}  \\
DDDM-VC      &$77.46 \pm 2.07$                             & 0.7                              & 2.87  & 3.84  & 3.21  \\
DDDM-VC + Ours   & \textbf{$78.61 \pm 1.66$}                            & \textbf{0.79} & \textbf{2.94 } & \textbf{3.91}  & \textbf{3.55} \\  \hline 
\end{tabular}
}
\vspace{-0.5cm}
\end{table}
\subsection{Ablation Study}
\begin{figure}[h]
  \centering
  \vspace{-0.3cm}
  \includegraphics[width=\linewidth]{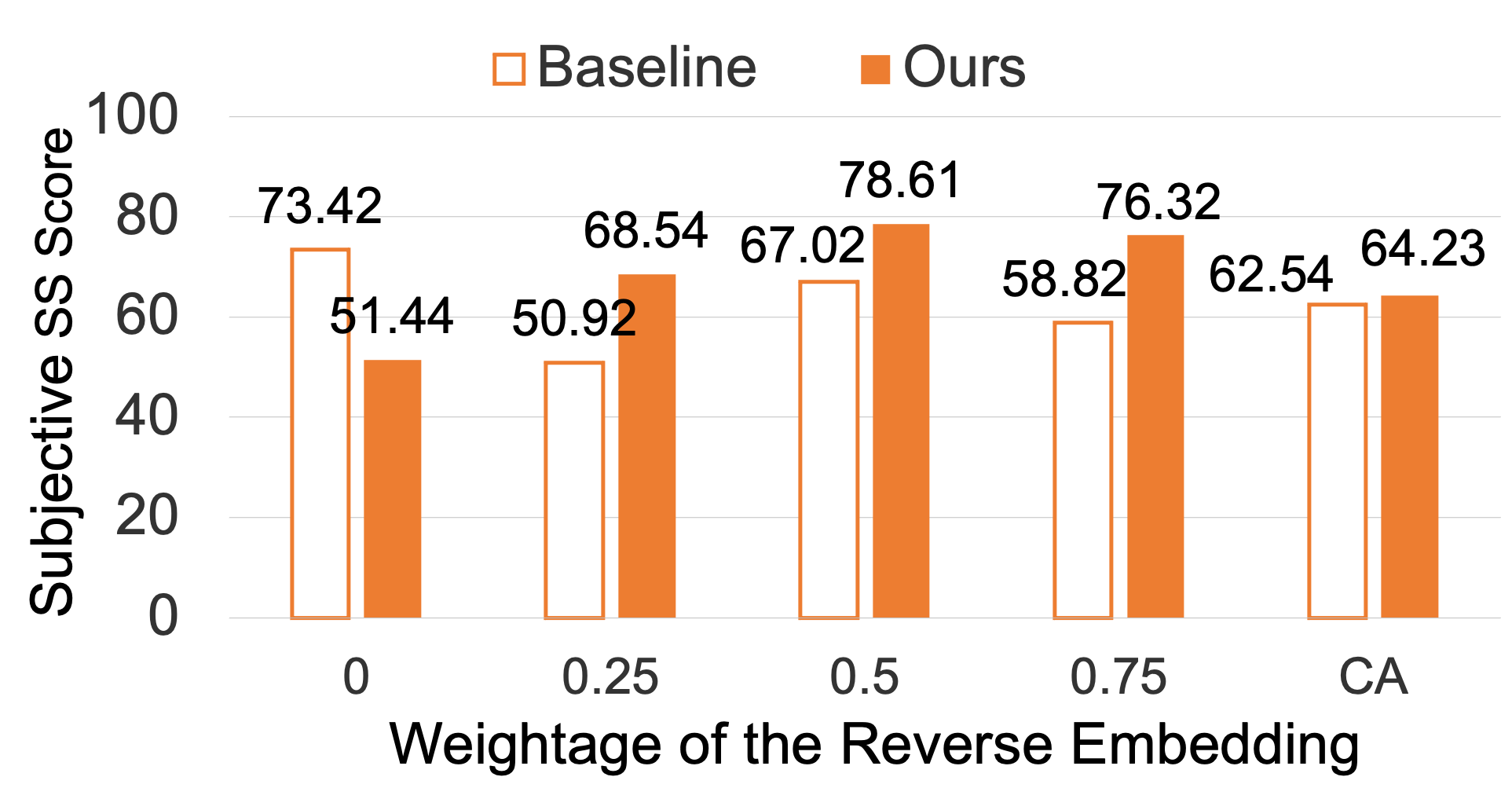}
  \vspace{-0.5cm}
  \caption{Ablation analysis for subjective speaker similarity scores w.r.t. Baseline(DiffVC)}
  \vspace{-0.5cm}
  \label{fig:ablation}
\end{figure}

Our ablation study explored the impact of combining the conventional speaker embedding with its time-reversed counterpart using different weighting schemes and the integration of cross-attention. Here, we again used MUSHRA-based subjective score to calculate speaker similarity from 25 subjects as mentioned in the previous subsection. We observed that an equal weighting of 0.50 for both embeddings consistently produced superior speaker similarity scores (as shown in Figure \ref{fig:ablation}).  This finding suggests that both the standard and time-reversed embeddings contribute equally valuable speaker-specific information, and the use of cross-attention further refines their integration, leading to enhanced overall performance in our voice conversion framework.
\section{Summary and Conclusions}
We leverage full‑utterance speech time reversal (STR) as a targeted data‑augmentation strategy within a diffusion‑based voice‑conversion pipeline to strengthen speaker embeddings. Whereas earlier efforts have applied reversal only to short segments or portions of the signal, our method inverts the entire utterance. This destroys intelligible linguistic content yet preserves global rhythmic and tonal contours, which, as our perceptual studies show, are alone sufficient to maintain speaker‑specific information. This finding supports the fusion of conventional speaker embeddings with those derived from time-reversed speech, providing a robust means to disentangle speaker identity from linguistic content. Our experimental evaluations on the LibriSpeech and VCTK databases, using a diffusion-based voice conversion framework, reveal that the proposed approach significantly improves speaker similarity scores while maintaining high speech quality. These results underscore the potential of STR to overcome data limitations in zero-shot VC scenarios and pave the way for future research to further refine and integrate such unconventional signal transformations into voice conversion systems.
\bibliographystyle{IEEEtran}
\bibliography{mybib}

\end{document}